\documentclass[twocolumn,prl,showpacs]{revtex4-1}

\usepackage{graphicx,amssymb,amsmath}
\usepackage{xcolor}
\usepackage[applemac]{inputenc}
\usepackage{epstopdf}

\begin{document}

\title{Nonlocal resonances in weak turbulence of gravity-capillary waves}
\author{Quentin Aubourg}
\author{Nicolas Mordant}
\email[]{nicolas.mordant@ujf-grenoble.fr}
\affiliation{Universit\'e Grenoble Alpes, LEGI, CNRS  F-38000 Grenoble, France}
\affiliation{Institut Universitaire de France, 103, bd Saint Michel, F-75005 Paris, France}

%\pacs{46.40.-f,62.30.+d,05.45.-a}

\begin{abstract}
We report a laboratory  investigation of weak turbulence of water surface waves in the gravity-capillary crossover. By using time-space resolved profilometry and a bicoherence analysis, we observe that the nonlinear processes involve 3-wave resonant interactions. By studying the solutions of the resonance conditions we show that the nonlinear interaction is dominantly 1D and involves collinear wave vectors. Furthermore taking into account the spectral widening due to weak nonlinearity explains that nonlocal interactions are possible between a gravity wave and high frequency capillary ones. We observe also that nonlinear 3-wave coupling is possible among gravity waves and we raise the question of the relevance of this mechanism for oceanic waves.
\end{abstract}

\maketitle

A large ensemble of nonlinear waves can exchange energy and develop a turbulent state. The statistic properties of such wave turbulence have been described theoretically for weak nonlinearity in the framework of the Weak Turbulence Theory (WTT). In this theory, only resonant waves are able to exchange significant amounts of energy over long times due to the weak nonlinear coupling. The predicted phenomenology of the stationary statistical states resembles that of fluid turbulence: energy is injected at large scales and cascades down scale to wavelengths at which dissipation takes over and absorbs energy into heat. A major difference with fluid turbulence is that analytical predictions for the stationary spectra (and other statistical quantities) can be derived for weak wave turbulence~\cite{R1,R2,R3}. Sea surface waves are among the pioneering physical systems that led to the development of the theory~\cite{R9}. The theory was applied to a vast amount of waves (in plasmas~\cite{R1}, solar winds~\cite{R4}, nonlinear optics~\cite{R7}, quantum superfluid vortices~\cite{R16}, vibrated elastic plates~\cite{R18},...).

For isotropic systems, the predicted energy spectrum $E(k)$ has the following expression 
\begin{equation}
E(k)=CP^{1/(N-1)}/k^\alpha
\end{equation}
where $k=|\mathbf k|$ is the wavenumber, $P$ the energy flux, $C$ a dimensional constant that can be calculated and $\alpha$ the spectral exponent. $N$ is the number of waves taking part in the resonances. Usually $N-1$ corresponds to the order of the nonlinear coupling term of the wave equation ($N=3$ for quadratic nonlinearities, $N=4$ for cubic ones,...). The waves have then to satisfy the resonance conditions such as $\mathbf k_1=\mathbf k_2+\mathbf k_3$ and $\omega_1=\omega_2+\omega_3$ (for 3-wave interaction). In some cases, these resonances conditions do not have solutions. This is the case in particular for gravity waves at the surface of water. The dispersion relation (for infinite depth) is $\omega=\sqrt{gk}$ and its negative curvature does not allow for solutions of the resonance conditions for 3 waves. Thus the resonances are expected to involve 4 waves. At small wavelengths, water waves are capillary waves for which the dispersion relation is $\omega=\left(\frac{\gamma}{\rho}\right)^{1/2}k^{3/2}$ whose curvature allows for 3-wave resonances ($\gamma$ is the surface tension and $\rho$ the density). The predicted spectra for water waves are thus expected to be $E(k)\propto P^{1/2}k^{-7/4}$ for capillary waves and $E(k)\propto P^{1/3}k^{-5/2}$~\cite{R2} for gravity waves. Laboratory experiments largely fail to reproduce this predictions, in particular for the gravity waves. In large or small waves tanks, the spectral exponent of the gravity waves is seen to vary strongly with the forcing intensity and to be close to the WTT predictions at the highest forcing magnitude, at odds with the weak nonlinearity hypothesis~\cite{R10,R11}. Furthermore the measurement of the exponent of the injected power is also different from that predicted by WTT~\cite{R11}. Recent work on water waves and vibrated plates suggest that wideband dissipation is most likely responsible for the latter observation~\cite{Humbert,R23,Deike}. Another experiment also suggest that several regimes of wave turbulence of water wave may exist depending on the intensity and frequency of the forcing~\cite{CoPr11}. Nevertheless the question of the order of the interaction remains of prime importance to test the theory. In this letter, we report a high order statistical analysis that directly probe the nonlinear interaction among waves. We implement a time resolved, 2D profilometry of the water surface deformation. The accessible wave lengths corresponds to capillary waves and to the gravity-capillary crossover at which the order of interaction supposedly switches from 3-wave at small wavelength to 4-wave at large wave lengths. We investigate to resonant interaction and the impact of the wave amplitude on the energy transfers.

%For small size experiment (ref) or for larger on (ref) the spectrum observed is far from the WWT prediction for weak forcing.In this letter we will focus on he gravity-capillary waves described by the following linear dispersion relation.
%\begin{equation}
%\omega^2=gk+\frac{\sigma}{\rho}k^3
%\label{equ1}
%\end{equation}
%Where $\sigma$ is the surface tension of water, $g$ the gravity and $\rho$ the density. We have performed water waves experiments using 3D measurements method and then computed waves correlation in order to test the WTT. We will see how resonant-waves interactions might be modified by the gravity-capillary cross-over.
\begin{figure}[!htb]
\includegraphics[width=6cm]{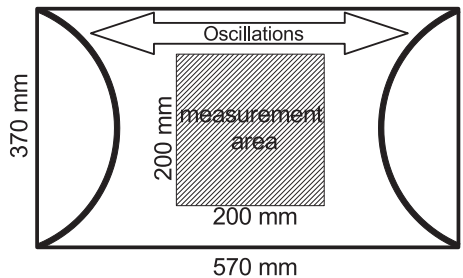}
\caption{Sketch of the experiment (Top view). The vessel is translating horizontally with a constant amplitude frequency modulated oscillation, which frequency is chosen in the interval $[0.5,1.5]$~Hz. Two curved walls are placed on left and right sides to improve the isotropy by diverging waves reflexion}
\label{fig0}
\end{figure}
The experimental setup consist of a rectangular plastic vessel of $57\times37$~cm$^2$ filled with 10~l of water to a rest height $h_0=5$~cm (fig.~\ref{fig0}). Two curved walls are used to improve the isotropy by divergent waves reflexion. Surfaces waves are excited by horizontally vibrating the vessel at frequencies in the range $[0.5,1.5]$~Hz. Waves are measured using the Fourier transform profilometry technique which enables a full space-time characterization of the waves~\cite{Cobelli1}. We used filtered softened water and thoroughly washed the tank. Similarly to Przadka {\it et al.}~\cite{Przadka2} we use anatase titanium dioxide particles (Kronos 1001) that do not alter the measured water surface tension and do not induce additional dissipation at the surface. Thanks to this pigment, it is possible to project a pattern at the very surface of water. When the water surface is deformed, the pattern seen by a camera is changed. The alteration of the pattern can then be inverted to recover the deformation of the surface~\cite{R22}. Here, the deformation of the pattern is recorded by a high-speed camera over a $20\times 20$~cm$^2$ surface at the center of the tank, with $1024\times1024$ pixels resolution at 250~frames/s. Datasets are made of 15 movies which duration is 87~s for each movie. 
%The vertical sensitivity has been estimated to 200$µm$ (? verifier) and the limitation of accessible wave number due to the phase demodulation is set by the period of the fringe pattern ($k<2\pi/L$).

\begin{figure}[!htb]
\includegraphics[width=7cm]{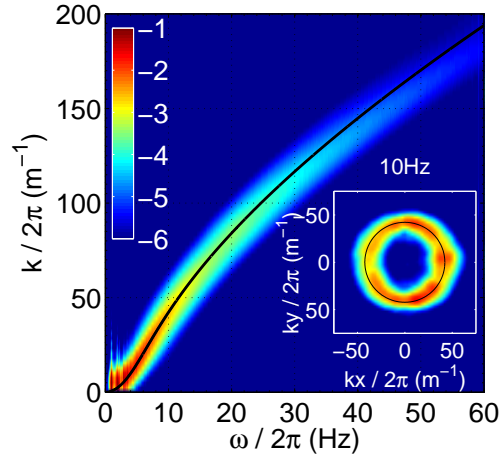}
\caption{Space-time Fourier spectrum of the velocity field of the waves $E^v(k,\omega)$ (see text for definition). The color scale is $\log_{10}E^v(k,\omega)$. The solid black line is the theoretical deep water linear dispersion relation for pure water $\omega^2=gk+\frac{\gamma}{\rho}k^3$ with $\gamma=72$~mN/m. Energy is localized on the dispersion relation and can be observed for frequencies up to 60~Hz. The crossover between gravity and capillary waves occurs at $k_c=\sqrt{\rho g/\gamma}=120 \pi$ corresponding to a wavelength of 1.7~cm and a frequency equal to 13~Hz. Inset: $E(\mathbf k,\omega)$ at $\omega/2\pi=10$~Hz. The energy distribution is fairly isotropic. The black circle corresponds to the linear dispersion relation. Color: $\log_{10}E(\mathbf k,\omega)$ between $-8$ and $-4$.}
\label{fig1}
\end{figure}
We show in fig.~\ref{fig1} the space-time power spectrum $E^v(k,\omega)$ of the velocity field $v=\frac{\partial \eta}{\partial t}$ where $\eta(x,y,t)$ is the altitude of the water surface. We first compute 
\begin{equation}
E^v(\mathbf k,\omega)=\langle |v(\mathbf k,\omega)|^2\rangle
\end{equation}
where $v(\mathbf k,\omega)$ is the space and time Fourier transform of the velocity. The time Fourier transform is computed by selecting a $16$~s time window of the signal. The average $\langle...\rangle$ is a time average over the time windows. $E^v(\mathbf k,\omega)$ is then summed over directions of the 2D wavevector $\mathbf k$ to provide a 2D picture of $E^v(k,\omega)$ (fig.~\ref{fig1}). Energy is seen to be concentrated around the linear dispersion relation of gravity-capillary waves
\begin{equation}
\omega=\left(gk+\frac{\gamma}{\rho} k^3\right)^{1/2}
\label{eq1}
\end{equation}
The isotropy of $E^v(\mathbf k,\omega)$ is shown in the inset of fig.~\ref{fig1} at a given frequency of 10~Hz. Energy is convincingly distributed over all directions. The energy concentration around the dispersion relation is due to nonlinear spectral widening predicted by the WTT framework. Note that no secondary branches of the dispersion relation are seen contrary to what was reported in~\cite{Herbert}. Our regime corresponds to the second regime of turbulence reported in~\cite{CoPr11} at the weakest magnitude of the waves. The wave steepness of our data is indeed small: $\sigma=\left\langle \sqrt{\frac{1}{S}\int_S \left\|\nabla h(x,y,t)\right\|^2dxdy} \right\rangle =0.025$ thus our wave field is weakly nonlinear.

\begin{figure}[!htb]
\includegraphics[width=8cm]{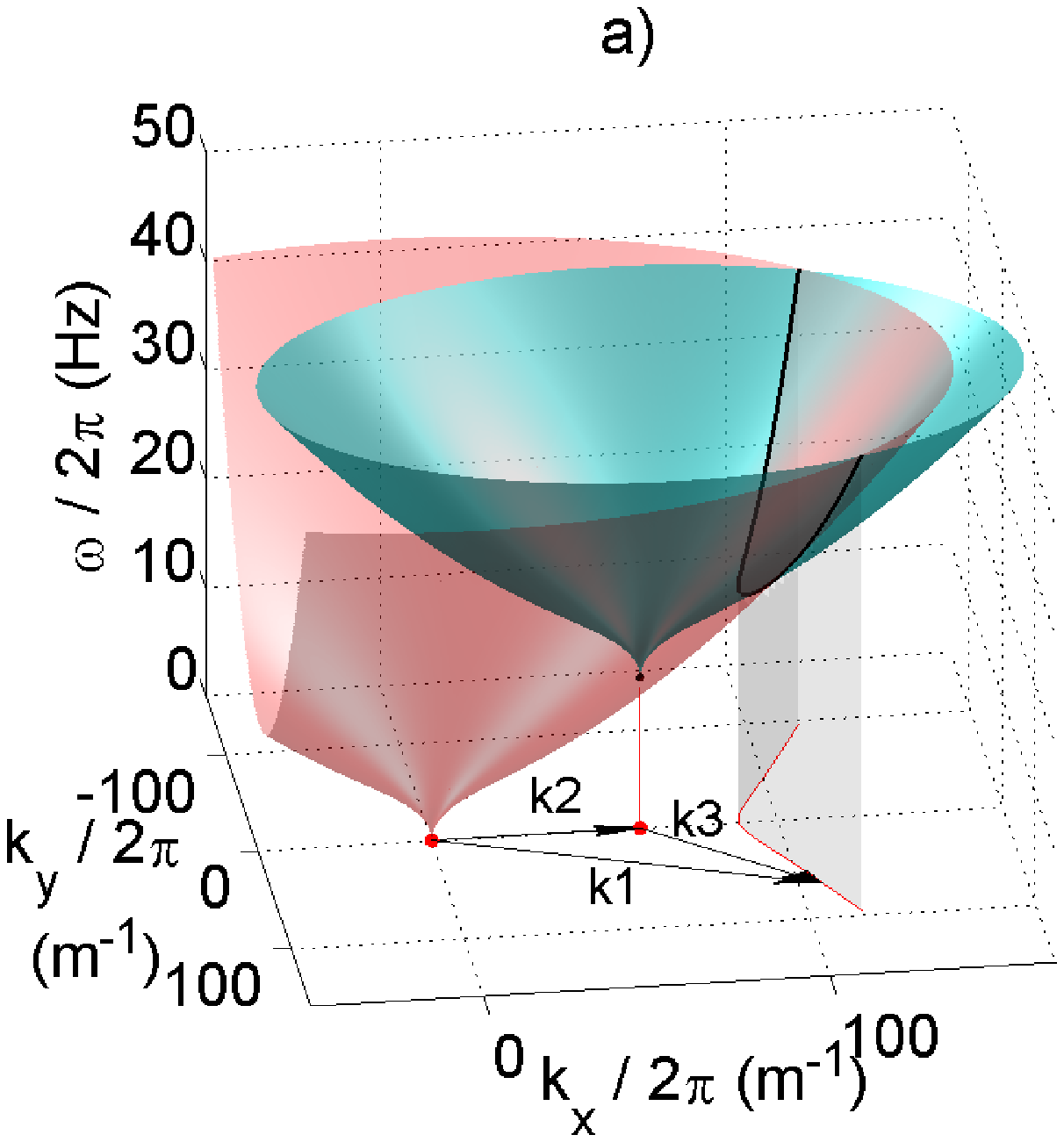}
\includegraphics[width=8.5cm]{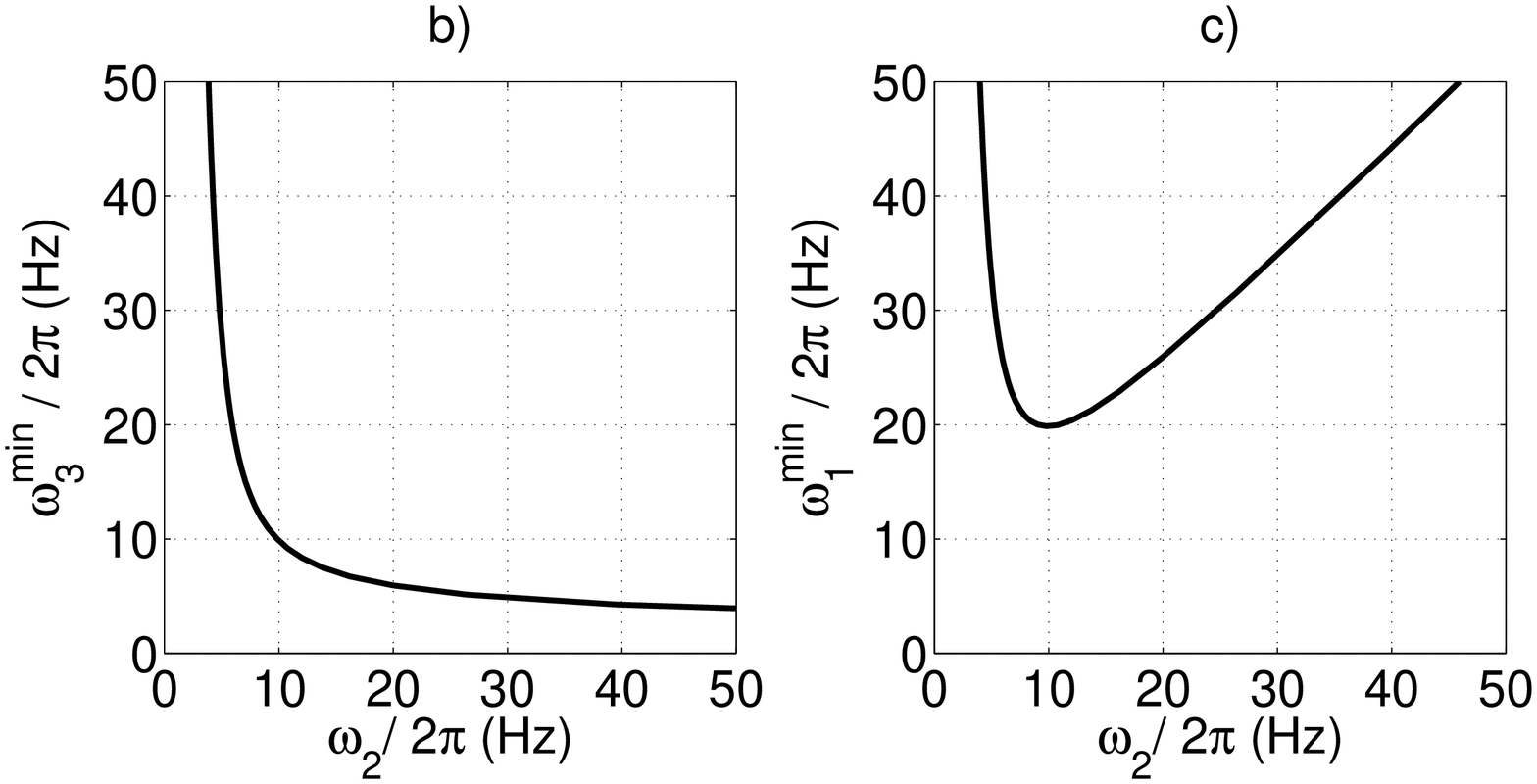}
\caption{Solutions for the 3 wave resonances of gravity capillary waves. (a) The red surface corresponds to the dispersion relation (\ref{eq1}). First a given value of $\mathbf k_2$ is chosen. The blue surface corresponds then to the sum $\omega(\mathbf k_2)+\omega(\mathbf k_3)$. The resonant wave at $\mathbf k_1$ must verify the resonance conditions $\mathbf k_1=\mathbf k_2+\mathbf k_3$ and $\omega_1=\omega_2+\omega_3$ and thus lie on the intersection of the red and blue surfaces (thick black line). The red line at the bottom of the axes corresponds to the projection of the black line. (b) Evolution of $\omega_3^{min}(\omega_2)$ that is the minimum value of $\omega_3$ for which there is a possible solution for a given value of $\omega_2$. (c) corresponding value $\omega_1^{min}$.}
\label{fig2}
\end{figure}
As discussed in the introduction, the order of nonlinear wave interaction depends on the order of the nonlinearity and the possibility or not to have solutions for the resonance equations. McGoldrick and then Simmons~\cite{Mcgoldrick1965,Simmons1969} has investigated the 3-waves resonant solutions in gravity-capillary regime. A 3D representation of the solutions for a given wavevector  $\mathbf k_2$ is shown in fig.~\ref{fig2}. As the curvature of the dispersion relation changes sign, solutions of the 3-wave resonance conditions exist even if the wave vector $\mathbf k_2$ lies in the gravity range. Such a solution exist only if the norm of the vector $\mathbf k_3$ exceeds a minimum value as can be seen from the red line in fig.~\ref{fig2}. This minimum value of $k_3$ is reached when the three wave vectors $\mathbf k_1$, $\mathbf k_2$ and $\mathbf k_3$ are collinear (1D situation). This can be translated in the frequency space: at a given frequency $\omega_2$, there is a minimum value of $\omega_3$ that allows for a resonant wave at $\omega_1$. In the following, we note $\omega_3^{min}(\omega_2)$ this minimum value (fig.~\ref{fig2}(b)) and $\omega_1^{min}(\omega_2)$ the corresponding value of $\omega_1$ (fig.~\ref{fig2}(c)). These values of the resonant frequencies correspond to 1D wave interaction. By scanning values of $\omega_2$, one observes that there is a overall minimum value for resonant $\omega_1^{min}$ corresponding to the degenerate case of Wilton wave : $\omega_1=2\omega_2=2\omega_3=2\pi\times19.6$~Hz (fig.~\ref{fig2}(c))~\cite{wilton1915lxxii}. Measurements performed by Hammack {\it et al.} ~\cite{Henderson1987} confirm the existence of these 3-waves resonant coupling.

\begin{figure}[!htb]
\includegraphics[width=8cm]{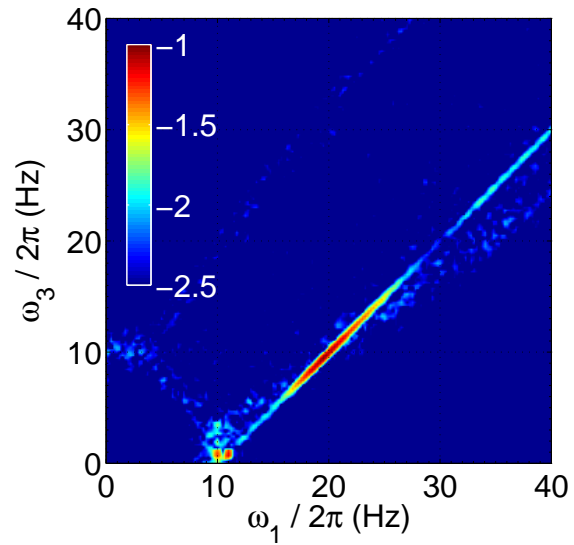}
\caption{Representation of 3-wave coherence $|C(\omega_1,\omega_2,\omega_3)|$ (see text for definitions). The specular light blue pattern on the background corresponds to statistical convergence noise so that the statistical convergence is about $10^{-2,5}$. Here $\omega/2\pi$ is chosen equal to 10~Hz. Thus the observed line of correlation lies above the convergence level by more than one order of magnitude at the maximum of correlation. This is the sign of the presence of significant 3-wave coupling in the signal. The line lies on the resonance curve $\omega_1=\omega_2+\omega_3$ as expected from such statistical estimator. Color is $\log_{10} |C|$.}
\label{fig3}
\end{figure}
To investigate the 3-wave coupling in our experiment, we study third order correlations of the velocity field. From $v(x,y,t)$, we compute the Fourier transform in time over $4$~s time windows so that to obtain $v(x,y,\omega)$. Correlations are then computed as
\begin{equation}
C(\omega_1,\omega_2,\omega_3)=\frac{|\langle\langle v^\star(x,y,\omega_1) v(x,y,\omega_2)v(x,y,\omega_3)\rangle\rangle |}
{\left[E^v(\omega_1)E^v(\omega_2)E^v(\omega_3)\right]^{1/2}}
\end{equation}
where $^\star$ stands for complex conjugation and the average $\langle\langle ...\rangle\rangle$ stands for an average over the time windows and a space average over $(x,y)$ positions on the image. $E^v(\omega)=\langle\langle |v(x,y,\omega)|^2\rangle\rangle$ is the frequency spectrum. With such a normalization, the coherence lies between 0 (no correlation) and 1 (perfect correlation). For a stationary (in time) signal such a third order correlation is expected to be nonzero only along the resonance line $\omega_1=\omega_2+\omega_3$. In order to check whether 3-wave correlation is truly observed, we show the coherence map at a given frequency (${\omega2}/{2\pi}=10$~Hz) in fig.~\ref{fig3}. A line of correlation emerges from the statistical convergence noise that confirms that 3-wave resonant processes are indeed present in the signal. Such a line can be observed at all values of $\omega_2$. Note that the correlation level is not homogeneous along the line showing preferential interaction among the waves. 
In order to look deeper in the nonlinear dynamics, we focus in the following on the bicoherence, defined as
\begin{equation}
B(\omega_2,\omega_3)=C(\omega_2+\omega_3,\omega_2,\omega_3)
\label{equ3}
\end{equation}
It corresponds to the extraction of the coherence observed on the resonant line of fig.~\ref{fig3} which has been checked to be significant and above the convergence noise level.

\begin{figure}[!htb]
\includegraphics[width=7cm]{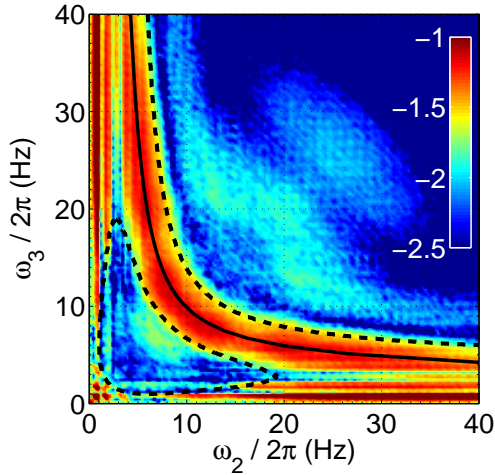}
\caption{Bicoherence $B(\omega_2,\omega_3)=C(\omega_2+\omega_3,\omega_2,\omega_3)$. The solid black line corresponds to the minimum value $\omega_3^{min}(\omega_2)$ allowing exact resonances (see text). The dashed line corresponds to the transformation of the solid line when allowing for a $\delta k/2\pi=5$~m$^{-1}$ uncertainty on the resonance condition (see text).}
\label{fig4}
\end{figure}
Figure~\ref{fig4} displays the bicoherence map corresping to the extraction of the coherence for all values of $\omega_2$. A noticeable organisation of the coherence is clearly visible. First a wide curved line of high coherence (about $10^{-1}$) is observed. The crest of coherence lies on the curve $\omega_3^{min}$ (solid black line) corresponding to unidirectional interaction of the waves as explained above. This observation suggests that the core of the nonlinear interaction in our experiment is quasi 1D. The isotropy of the space-time spectrum of fig.~1 may be somewhat artificial and mostly due to chaotic mixing of the waves due to the boundary conditions (reflexion of curved walls) rather than nonlinear directional spreading. 

Another region of high coherence is seen also for weak values of $\omega_3$ and large values of $\omega_2$ (bottom-right part of fig.~\ref{fig4} along the axis). This region is made of two horizontal lines of high correlation values (a symmetric region exists next to the vertical axis). The discrete character of these two lines suggests that it corresponds to interactions of capillary waves with the first modes of our finite domain. Note that the eigenfrequencies of these modes lie in the gravity domain ($0.9$~Hz, \& $1.7$~Hz). The existence of these regions seems paradoxical as no exact resonant interaction of waves is possible below the curve $\omega_3^{min}$. 

To explain this paradox, let us note that the coherence line along $\omega_3^{min}$ has a finite (nonzero) width. This width may be associated to the similar width of the energy concentration along the dispersion relation of fig.~\ref{fig1}. This width is due to nonlinear effects: indeed the coherence of the linear waves is altered by the nonlinear energy exchanges with the other waves~\cite{R2,R21}. The width of the dispersion relation is thus a measure of the degree of nonlinearity of the system. Let us assume that $\delta k$ is the width in wavenumber of the dispersion relation ($\delta k$ is taken constant for simplicity). It can be interpreted as a nonlinear uncertainty on the determination of the wavenumber. Thus the resonance condition translates into an inequality $|\mathbf k_1-\mathbf k_2-\mathbf k_3|<\delta k$ rather than an equality to zero. For simplicity, let us assume that as suggested by the data in fig.~\ref{fig4} the nonlinear coupling occurs solely through a 1D mechanism. We look for the solutions $k_1=k_2+k_3\pm\delta k$ and $\omega_1=\omega_2+\omega_3$. The two corresponding solutions are shown as the dashed lines in fig.~\ref{fig4} for $\delta k/2\pi=5$~m$^{-1}$. This value of $\delta k$ is reasonable in view of the width of the dispersion relation in fig.~\ref{fig1}. The two dashed lines spectacularly encircle the blue regions of very low coherence and highlight the regions of high coherence. The new region of high coherence delimited by the dashed lines incorporates all previously discussed regions. This observation removes the apparent paradox. When one takes into account the nonlinear spectral widening, the possible interactions are much more numerous, including frequencies well below the $\omega_3^{min}$ curve. In particular it opens a region of strongly nonlocal interaction of a very low frequency gravity mode and 2 much higher capillary waves. For example, the mode at 0.9 Hz can interact with the whole interval of frequencies and thus initiate the energy cascade. 

Furthermore, as can be seen in the bottom left corner of the picture, very low frequency modes can also interact among each other. As discussed in the introduction, the nonlinear interaction of gravity modes is usually assumed to involve 4 waves but incorporating a small nonlinear spectral widening may allow for 3-wave quasi resonances that may actually be more efficient to transfer energy. This possible interaction remains to be quantified precisely as in our simplified calculation, the widening is taken constant whereas it depends usually on the frequency~\cite{R21}. If confirmed at larger scale, this mechanism could also be responsible for the discrepancy of the observed spectral exponents in the gravity range as compared to the WTT predictions that assume 4-wave interactions.

In conclusion, we have shown that weak turbulence of water waves near the gravity-capillary crossover relies on a quasi 1D 3-wave interaction. The theoretical reason for the selection of this 1D mechanism while 2D resonances are {\it a priori} allowed remains to be investigated. Taking into account the nonlinear spectral widening into the study of the resonance conditions appear to be of prime importance at it changes significantly the range of possible interactions. In particular it allows for strongly nonlocal interactions between gravity and capillary waves and for 3-wave interaction among gravity waves, previously assumed to be impossible.

\begin{acknowledgments}
We thank Kronos Worldwide, Inc. for kindly providing us with the titanium oxide pigment. We acknowledge many discussions with M. Berhanu, L. Deike and E. Falcon.
\end{acknowledgments}
\bibliography{biblio7}

 \end{document}